\documentstyle[editedvolume,numreferences]{crckapb}
\input psfig.sty

\def\kms{{km\thinspace s$^{-1}$}}
\def\13co{$^{13}$CO}
\def\h2{H$_2$}

\def\deg{\ifmmode^\circ\else$^\circ$\fi}
\def\dege{{\ifmmode^\circ\else$^\circ$\fi}\ }
\def\solar{\ifmmode _{\mathord\odot}\else$_{\mathord\odot}$\fi}
\def\sun{$_\odot$}
\def\sune{{$_\odot$}\ }

\def\msun{M\sun}

\def\msune{{M\sun}\ }

\def\hper{\ifmmode \rlap.^{h} \else $\rlap{.}^h$\fi}
\def\mper{\ifmmode \rlap.^{m} \else $\rlap{.}^m$\fi}
\def\sper{\ifmmode \rlap.^{s} \else $\rlap{.}^s$ \fi}
\def\degper{\ifmmode \rlap.^{\circ} \else $\rlap{.}^{\circ} $\fi}
\def\arcmper{\ifmmode \rlap.{' } \else $\rlap{.}' $\fi}
\def\arcsper{\ifmmode \rlap.{'' } \else $\rlap{.}'' $\fi}
\def\cc{cm$^{-3}$}

\def\c2{cm$^{-2}$}
\def\2e{$^{2}$\ }
\def\2{$^{2}$}
\def\3e{$^{3}$\ }
\def\3{$^{3}$}
\def\-2e{$^{-2}$\ }
\def\-2{$^{-2}$}
\def\-3e{$^{-3}$\ }
\def\-3{$^{-3}$}

\def\m12{\magnification=1200}

\def\deg{\ifmmode^\circ\else$^\circ$\fi}              
\def\arcs{\ifmmode {'' }\else $'' $\fi}           
\def\arcm{\ifmmode {' }\else $' $\fi}             

\def\oversim#1#2{\lower0.5ex\vbox{\baselineskip=0pt\lineskip=0.2ex
     \ialign{$\mathsurround=0pt #1\hfil##\hfil$\crcr#2\crcr\sim\crcr}}}
\def\simgt{\mathrel{\mathpalette\oversim>}}
\def\simlt{\mathrel{\mathpalette\oversim<}}
\def\thirteenco   {$^{13}{\rm CO}$}
\def\nhtwo  {n_{\rm H_2}}
\def\calt{{\cal T}}
\def\calw{{\cal W}}
\def\fdg{\hbox{$.\!\!^\circ$}}

\begin{opening}
\title{MOLECULAR CLOUDS}
\author{LEO BLITZ}
\institute{Astronomy Department, University of California\\
           Berkeley, CA USA\\
	       blitz@gmc.berkeley.edu}

\author{JONATHAN P. WILLIAMS}
\institute{National Radio Astronomy Observatory\\
           Tucson, AZ  USA\\
           jpwilliams@nrao.edu}
\end{opening}
\runningtitle{MOLECULAR CLOUDS}
\begin{document}


\section{Introduction}

	All known star formation is thought to occur in molecular
clouds.  The association of molecular clouds with star formation is so
strong that it is generally assumed that wherever there are young
stars, one will always be able to find molecular gas, even when there
is evidence to the contrary
(e.g. the TW Hya association; Rucinski \& Krautter 1983). Indeed,
when sufficiently sensitive observations are made, the association of
star formation with molecular clouds is observed in all environments,
galactic or extragalactic.  Evidence for star formation in molecular
clouds has even been observed as far back as z = 4.7 (Omont {\it et
al.} 1996), and sensitive new instruments such as the Millimeter Array
and the Square Kilometer Array should make it possible to detect star
forming molecular clouds at the earliest times.

	The general goal of molecular cloud studies is to determine
how the interstellar medium produces molecular clouds, especially the
Giant Molecular Clouds (GMCs; M $> 10^4$ \msun) which are responsible
for the vast majority of all star formation, and how the molecular
clouds in turn produce stars and clusters.  We take the view that
once
a core within a molecular cloud can no longer support itself against
gravity, beginning the inexorable process of collapse into a single
star or binary, the study of the molecular gas falls into a different
regime. The inexorability of the star formation process also 
occurs at the GMC level; there is only one GMC of dozens known 
within about 3 kpc of the Sun with
scant evidence of star formation (see \S3.3).

We wish to obtain answers to the four fundamental questions below:

1) How do GMCs form?

2) How do single stars (and binaries) form?

3) How do clusters form?

4) What determines the Initial Mass Function (IMF)?

	We will begin below by discussing some of the observational
progress that has been made in answering the first question.  We will
deal with the second question only in passing; it will be dealt with
extensively by articles in this volume by
Shu et al., Lada, Myers, and others.
We will address steps leading to the answers to the
third and fourth questions by looking for clues from the analysis of
the structure of molecular clouds.

	This review is meant primarily to address the progress made in
the study of molecular clouds since the first Crete meeting on star
formation, {\it The Physics of Star Formation and Early Evolution}
(Blitz 1991), which discusses work on the subject through 1990.  
Other useful reviews include the article in this volume by
McKee, which gives a good theoretical picture of the physics of
molecular clouds, and the article in {\it Protostars and Planets IV} by
Williams, Blitz \& McKee (1999) which discusses progress in molecular
cloud studies since {\it Protostars and Planets III} (Blitz 1993).

\section{Formation of Molecular Clouds}
\subsection {Galaxy Scale Issues}

	The physics of the formation of GMCs is one of the major unsolved
problems of the interstellar medium.  Although many papers have been 
written on the subject, especially in the late 1970s and early 1980s,
it is not yet known what the dominant formation mechanism is, or even
what the relative importance of gravity, radiation and magnetic fields
are in the cloud formation process.  For
example, GMCs are known to be self-gravitating because their mean
internal pressures exceed that of the general ISM by about an order of
magnitude (e.g. Blitz 1991).  Since they are known to be at
least as old as the oldest stars identified to have formed from them
(e.g. $\sim$20 My in the case of Orion; Blaauw 1964), they must be
stable for at least that long.  It is then reasonable to conclude that
gravity is one of the key elements in the formation of
self-gravitating, relatively stable clouds. However,  molecular
clouds found at high galactic latitude ({\it high latitude clouds}; HLCs)
also generally  have turbulent
pressures greater than the mean mid-plane ISM pressure,
but have masses a few orders of magnitude smaller than the GMCs. 
Typically they are
far from being self-gravitating (Magnani, Blitz \& Mundy 1985; Reach,
Wall \& Odegard 1998) and gravity cannot have been a factor in their
formation.  Is gravity important only in forming clouds 
that exceed a certain minimum mass?

	Mechanisms proposed for molecular cloud formation can be
divided into three general categories: collisional agglomeration of
smaller clouds (e.g. Kwan 1979; Scoville \& Hersh 1979; Stark 1979;
Cowie 1980; Kwan \& Valdez 1983), gravi-thermal instability (e.g.
Parker 1966; Mouschovias, Shu \& Woodward 1974; Shu 1974;
Elmegreen 1982a,b), and the pressurized accumulation in shocks, either
in supernovae (\"Opik 1954; Herbst \& Assousa 1977) or in Galactic shocks
(e.g. Woodward 1976).  These mechanisms were reviewed by Elmegreen
(1990; and references therein) who concluded that the most likely
formation 
mechanism is the gravi-thermal instability applied to the {\it cloudy}
ISM [his italics].  However, direct evidence has been very hard to come
by and it has not been possible to apply the ideas to other galaxies
or even to other parts of the Milky Way such as the Galactic
Center.  There may be clues in the angular momentum distribution
of GMCs (Blitz 1993) but this has received little attention; it is
not known, for example, whether GMCs in the disk rotate faster than
they do in the solar vicinity, as expected from gravitationally formed
clouds in a differentially rotating disk, or whether the
counter-rotating clouds in the local neighborhood are pathological, or
common throughout the disk. In one of the few papers on the subject,
Phillips (1999) has analyzed
the role of angular momentum in GMC support and formation:
isolated clouds tend have angular velocity vectors perpendicular to
the Galactic plane suggesting that their spin arises from Galactic
shear, but internal structures have more randomly oriented spin
axes due, possibly to turbulence, or dynamical interactions.

\begin{figure}[htb]
\centerline{\psfig{figure=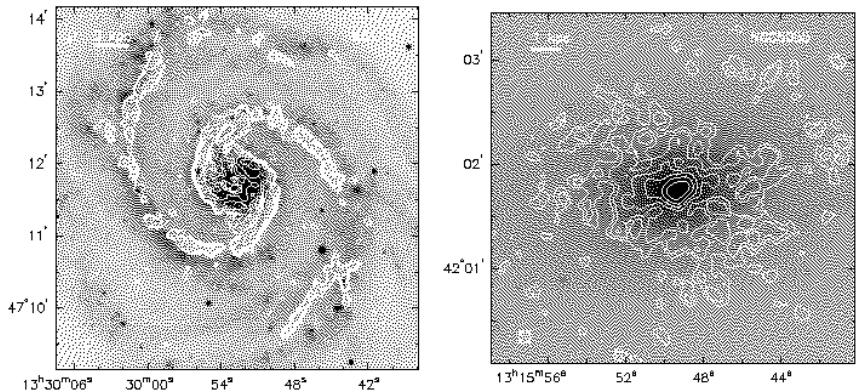,height=2.2in,angle=-90,silent=1}}
\caption{{\it Left}: CO in M51 from the BIMA Survey of Nearby Galaxies
(SONG), overlaid on an optical image of the galaxy.  The image
contains zero spacing data and thus samples all of the spatial
frequencies to the resolution limit (shown by the small box in the
lower left). The CO is very strongly concentrated to the spiral arms
and lies upstream of most of the ionized gas. {\it Right}: BIMA SONG
image of the CO in NGC 5055 overlaid on an optical image showing that
the galaxy is devoid of large-scale spiral structure at visible
wavelengths.  Some of the off-nuclear CO peaks are associated with weak
spiral arms seen in the near infrared. The difference in CO morphology
between these two galaxies is rather striking.}
\end{figure}

	Nevertheless, two questions regarding molecular clouds
which give us important clues for how GMCs form, have been settled
in the last decade:  1) GMCs have lifetimes of 2 $\times 10^7 < \tau < 
1 \times 10^8$y, considerably shorter than a Galactic rotation
period; 2) In galaxies with strong, well-defined spiral arms,  molecular
clouds are generally confined to the arms.  

	The question of cloud lifetimes, while hotly debated in the
early 1980s, seems to have been settled by two sets of observations.
The first is a good calibration of the \h2/HI surface density ratio,
$\Sigma$(\h2)/$\Sigma$(HI), everywhere in the Milky Way (Dame 1993).
This result coupled with a downward revision of the I(CO)/N(\h2) ratio
from the EGRET experiment on GRO (Hunter et al. 1997), implies that
there is no radius in the Milky Way where $\Sigma$(\h2)/$\Sigma$(HI) is
significantly greater than 1.  Thus the mass flow arguments suggesting
long cloud
lifetimes ($\simgt 10^9$ y; e.g. Solomon and Sanders 1980) are no
longer applicable in the disk
(see however \S2.3 below for a discussion of the Galactic Center).  The 
second is that the depletion time for the
molecular gas due to star formation over the entire Galaxy is about
$2-5\times$ 10$^8$~yr independent of radius
(Lacey \& Fall 1985; Blitz 1995), setting a strict
upper limit to the lifetime of the molecular gas (rather than just the
molecular clouds) in the Galaxy.

	The degree of confinement of the molecular gas to the spiral arms has
long been understood to be a direct test of GMC lifetimes, and the work
of Cohen et al. (1980) has recently been improved with more sensitive data
from the same telescope by Digel et al. (1996) and the new
FCRAO outer Galaxy CO survey (Heyer \& Terebey 1998).
The CO integrated intensity contrast between the arm and interarm
regions in the outer Galaxy is at least a factor of 28, indicating
that the gas that enters a molecular arm is overwhelmingly atomic.
Some of the
results are discussed in greater detail below.  With the improvement in
resolution and sensitivity of the millimeter-wave interferometers it
has now become possible to determine the degree of confinement of the
molecular gas to the spiral arms in a wide variety of other galaxies.
Figure~1a shows an image of the CO emission from M51 with the BIMA
Array.  This image contains data from all spatial scales down to the
resolution limit and therefore contains all of the CO flux within the
region surveyed.  In M51, as in most of the other galaxies with strong
spiral arms in the BIMA galaxy survey, the CO is closely confined to
the spiral arms.  However, in NGC 5055 (Figure 1b), there is a much
smaller tendency for the CO to lie in spiral arms.  The optical image
shows no large scale spiral structure at all, though near infrared
images show weak underlying spiral arms (Thornley 1996).  Nevertheless
NGC 5055 does show evidence for spirality, short incoherent spiral arm
segments, that may play a role in GMC formation.  Based on H$\alpha$
images of the galaxy, It appears that the molecular clouds in NGC 5055
are not significantly different in their star forming properties from
those in the Milky Way (Thornley 1996), because the star formation rate
does not appear to be suppressed with respect to that measured in our
own Galaxy.

	We do not know, however, whether the physical properties of the
individual molecular clouds in NGC 5055 differ from those in the Milky
Way; the combination of sensitivity and resolution needed to find out will
probably require the MMA.  So, while the role of spiral arms in GMC
formation is amply demonstrated in M51, M100, the Milky Way (see
below) and other galaxies, it is unclear to what degree the spiral
arms are {\it necessary} for GMC formation.  Furthermore, there are
some grand design spirals such as M81 that are weak in CO, so much so
that it is not known to what degree the molecular clouds are confined to
the spiral arms.  More sensitive observations can settle the issue
in M81 with present instruments.

\subsection {The Chaff}

We now wish to address the question of GMC formation in the disk of
the Milky Way.  In particular, recent observational progress has made
it possible to address directly the role of molecular cloud
agglomeration in the formation of GMCs.

New sensitive surveys of the molecular gas in the second
Galactic quadrant with the FCRAO telescope 
(Heyer et al. 1998) and around one GMC 
(Mon OB1; Oliver, Masheder \& Thaddeus 1996) 
have shown that there is a great
deal of low surface brightness (i.e. low surface density) 
molecular gas in the spiral arms in the Galaxy and in the vicinity of GMCs.
This gas bears a striking resemblance to the HLCs discovered in the mid-1980s 
(Blitz, Magnani \& Mundy 1984).  We may ask, is this molecular
``chaff'' seen in the high sensitivity surveys the same as the HLCs and
can this molecular gas agglomerate to form the GMCs in the Milky Way?

	We first note that like the HLCs, most of the low column
density gas in the Oliver et al. study is not self-gravitating; the
dynamical masses typically exceed the luminous masses by more than an
order of magnitude.  The derived luminous masses of the clouds that
compose the Mon OB 1 chaff
are higher than the typical HLC masses, but this may be because Oliver
et al. artificially placed all the clouds at a distance of about 1 kpc even
though the clouds might be at any distance along the line of sight from
0 to about 1 -- 1.5 kpc. The Oliver et al. map bears a striking
resemblance to the FCRAO maps of Heyer et al. (1998), suggesting that
the chaff permeates the entire outer Galaxy but with a decreased
surface density between the arms.  An example of the FCRAO survey in
the range $102\deg < l < 142\deg$ is shown in Figure 2a.
A longitude-velocity plot over the same range is shown in Figure 2b.

\begin{figure}[htb]
\centerline{\psfig{figure=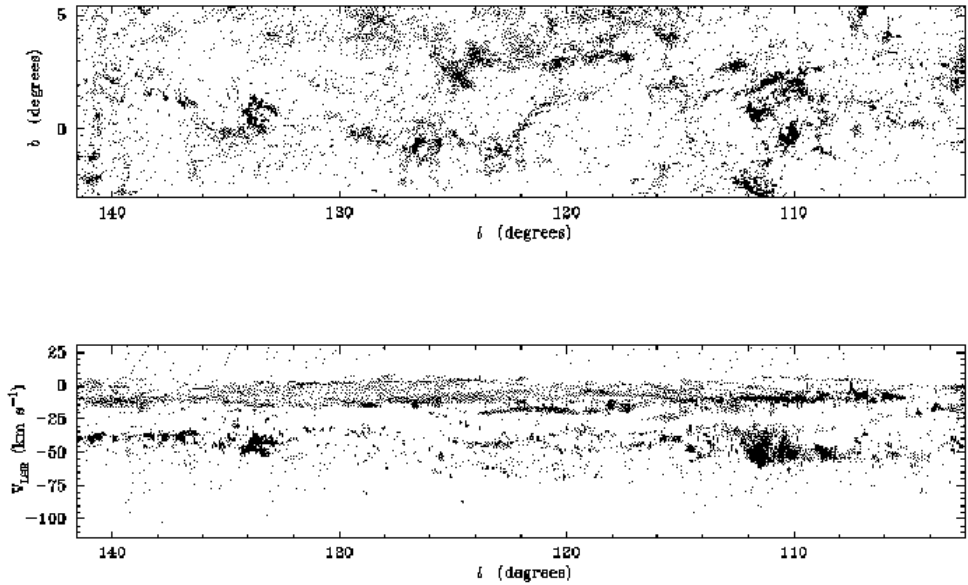,height=3.2in,angle=-90,silent=1}}
\caption{CO emission in the outer Galaxy (from Heyer et al. 1998).
Top panel is a velocity integrated $l-b$ plot, lower panel is a $l-v$
plot (integrated over the latitude range of the survey) showing the
local and Perseus spiral arms. Note the almost complete absence of
molecular gas between the two spiral arms at about 0 -- -10 \kms and -40
\kms.
The high spatial dynamic range of this survey shows the large scale
distribution of molecular gas in the ISM in exceptional detail.}
\end{figure}

	Heyer \& Terebey (1998) have determined the molecular cloud
spectrum ($dN/d\ln M$) for more than 1500 clouds in the Perseus Arm
and found a power law exponent of $-0.75$, similar to the value
$-0.6$ found for inner Galaxy clouds
(e.g. Sanders, Scoville, \& Solomon  1985). Furthermore, their
measurements included clouds with masses as low as $\sim$ 100 \msun,
well into the mass range of HLCs.  This measurement has several
implications.  First, it suggests that the chaff (M $< 10^3$ \msun)
is part of the same population as the self-gravitating
GMCs, and may very well share the same formation process:  there is no
obvious change in slope of the mass spectrum at the point where 
the clouds are no
longer self-gravitating.  Second, Heyer (1999) argues that, like the
low column density clouds near MonOB1, the
clouds that make up the chaff are not self-gravitating.
Third, the slope implies that as for GMCs,
most of the mass in chaff is contributed by the highest mass clouds.
Thus even though Oliver et al. have probably erred on the masses of
many individual clouds as the result of incorrect distance assumptions,
the total estimated mass is probably not too much in error.

	Oliver et al. (1996), find that the total mass of chaff near 
Mon OB1 is 3 $\times 10^4$ M\sun.  This is contained
within a Galactic surface area of about 1 -- 2 $\times 10^5$ pc\2, thus
the surface density of the chaff is 0.15 -- 0.3 M\sune pc$^{-2}$.  The
surface density of the HLCs is found to be 0.1 -- 0.2 M\sune pc$^{-2}$
(Magnani, Blitz \& Mundy 1985; Magnani, Lada, \& Blitz 1986; 
Reach et al. 1998; Magnani, Hartmann \&
Thaddeus 1999), close to the Oliver et al.
determination.  The HLCs thus have the same surface density as the
chaff, and they are each non-self-gravitating clouds of comparable
mass.  The chaff therefore appears to be the Galaxy-wide
identification of what were first identified as high-latitude
molecular clouds.

	Figure 2 suggests that we may extrapolate the Oliver et al.
values to the Galaxy as a whole. Assuming a constant surface density,
we find that the total mass of the chaff is 7.5 -- 15 $\times 10^7$
M\sun, and that the total mass in GMCs is $\sim 9 \times 10^8$ M\sune
(Dame 1993 -- scaled to the Hunter 1997 recalibration of I(CO)/N(\h2)).  
This difference implies that there is not enough mass in small
molecular clouds to form the GMCs through collisional agglomeration.
That is, if the number of GMCs is roughly in steady state and the
formation time equals the dissolution time, then, averaged over the
Galaxy, there would have to be approximately equal gas masses
in GMCs and in the chaff.
As Cowie (1980) showed, this need only to be true in the spiral arms, since
orbit crowding can enhance the surface density of small clouds in the
arms. However, the mass of chaff is well below that of the GMCs by
about an order of magnitude, even
in the spiral arms. It therefore appears that GMCs in the disk of the
Milky Way {\it must} form by condensation from the HI, rather than
from pre-existing molecular gas.

 	What does this imply for GMC formation and evolution?  First,
the short lifetime of molecular clouds implies that molecular gas
recycles through the ISM fairly rapidly, and that the instantaneous
recycling approximation often used in Galactic chemical evolution calculations
is a good one.  Second, it suggests that the collisional agglomeration
of pre-existing molecular clouds as a formation mechanism for GMCs
seems to be untenable in any form.  Third, although spiral shock
induced formation of GMCs seems to be indicated in many galaxies, it
does not explain why there is extensive molecular gas and star
formation in galaxies with weak or absent spiral arms.  The formation
of molecular clouds therefore seems to occur by condensation from the
HI in conjunction with some other mechanism.  This conclusion
has already been discussed previously (e.g. Elmegreen 1990, Blitz 1991,
1993); the radius from which the \h2 forms in the solar vicinity
is about 150 pc for a typical GMC with radius $\sim 30$~pc.
Although these conclusions seem secure within the
Galactic disk, and are probably also true even at the peak of the
molecular ring, the situation is dramatically different for GMCS near
the Galactic Center where the ratio $\Sigma$(\h2)/$\Sigma$(HI) $\simgt$
100 (e.g. Liszt \& Burton 1996).

\subsection{The Association of Atomic and Molecular Gas}

	HI envelopes around molecular clouds are quite common
(e.g., Moriarty-Schieven, Andersson \& Wannier 1997;
Williams \& Maddalena 1996).
Figure 3 shows an example of such a cloud around the Rosette Molecular
Cloud (Williams, Blitz, \& Stark 1995).
In the solar vicinity the mass of the molecular envelopes seem to be
about the same as the mass of molecular gas in a GMC (Blitz 1990),
but the atomic gas is considerably more extended. These envelopes may
be remnants of the atomic clouds that condensed to form the GMCs
or may be the photodissociated gas from the GMCs, since the
extinction is $\sim 0.25-0.5$ A$_{\rm V}$, just what is needed to
shield the CO from the interstellar UV field. Most probably, they are a
combination of both.
In the solar vicinity, the separation between the HI
envelopes is considerably larger than their diameters: the mean
distance between GMCs is about 500 pc, and the molecular cloud/HI
envelope complexes have diameters of about $150-200$ pc (Blitz 1990).  
The complexes are thus distinct from the background HI, indeed that
is how they are identified.

\begin{figure}[htb]
\centerline{\psfig{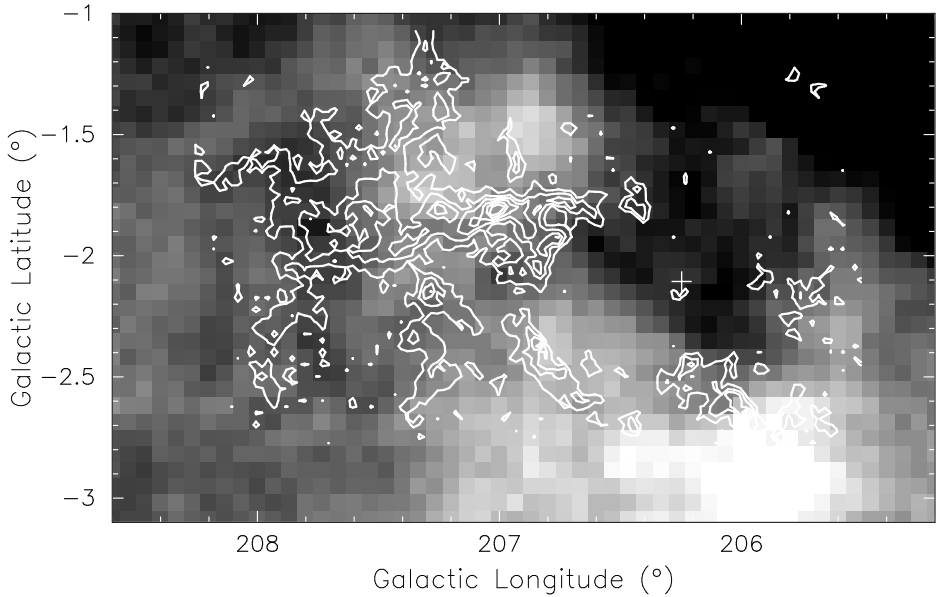}}
\caption{An HI envelope around the Rosette molecular cloud.
The grayscale, range 450 to 620~K~km~s$^{-1}$, shows
HI data from Arecibo observations by Kuchar \& Bania (1993).
Contours, beginning from and with increment 18~K~km~s$^{-1}$, are
CO emission from Bell Labs observations by Blitz \& Stark (1986).
Emission has been summed over a velocity range $v=4-25$~km~s$^{-1}$.
The cross marks the OB association that lies at the center of the
Rosette nebula and has cleared out the neutral gas.
The regions of strong HI emission (lighter colors)
lie on the CO cloud boundaries, forming an envelope around the cloud.}
\end{figure}

	In the inner Galaxy the situation is somewhat different. Near
the peak of the molecular ring, the surface density of molecular gas is
about equal to that of the atomic gas (Dame -- 1993 modified by the new
calibration of Hunter et al. 1997).  If the GMCs in the ring have atomic
envelopes with the HI mass equal to the \h2 mass,
as is true locally, then all of the atomic gas
in the molecular ring would be associated with GMC envelopes, leaving
little atomic gas for a true intercloud medium. This may
not be a problem, however: the molecular gas surface density is a factor
of $\sim$ 5 -- 6 greater in the molecular ring than locally so the mean
distance between GMCs $\sim 500/\sqrt{5}\simeq 200$~pc,
about the same as the diameter of the CO/HI
GMC complexes.  It may be that the envelopes
merge and form a general background within which the GMCs are
located.  This might explain why it has been more difficult to
associate atomic hydrogen clouds with GMCs in the molecular ring than
it is locally.

	Within several hundred parsecs of the Galactic center, the
situation is drastically different.  First, the gas is almost entirely
molecular; only about 1\% of the surface density of the gas is atomic
(e.g. Liszt \& Burton 1996).
In this region of the Galaxy, gas that becomes neutral must quickly
become molecular, and if the formation and dissolution of molecular
clouds goes through an atomic phase, that phase must be very brief.

The relation of the atomic to molecular gas must also be very different in
the central 300 pc of the Galaxy because of the large interstellar gas
pressure due to the deep stellar potential of
the bulge/bar.  In the bulge, the interstellar gas pressure is both
predicted and measured to be almost three orders of magnitude greater
than it is locally (Spergel \& Blitz 1992), and both the atomic and
molecular gas are highly overpressured compared to the solar
vicinity.
The HI is particularly problematic because at a pressure P/k $\simeq
10^7$ K\cc, and a maximum temperature of 10$^4$ K, the mean density of
the atomic gas near the center 
must be $\geq 10^3$ \cc, a density that is as high as
the typical molecular gas density in local GMCs.  How can this be?
Apparently, the HI can only exist as either very small dense knots, in
which case it would be difficult to avoid turning molecular, or
as thin, high density 
sheets on the surfaces of the molecular clouds at the center.
Because the molecular gas already has a low volume filling fraction, 
the ionized gas must make up almost all of the volume in the
bulge/bar.  This picture of the relationship between the atomic and
molecular gas has not been directly verified by observation, but a
combination of VLA observations with archival molecular cloud data should be
able to do so.  

The effect on star formation of this high pressure is hard to predict
without better knowledge of the power law $\gamma$ relating the
density and pressure of the gas.  For example, for a perfect gas,
the Jeans mass of clumps is 

$$M_J = 10 {v_{rms}^4 \over (G^3 P_0)^{1/2}} =
10^6 \left({v_{rms} \over 5 {\rm km s}^{-1}}\right)^4 \left({ P_0
\over 5 \times 10^6 K {\rm cm}^{-3} }\right)^{-1/2}
M_\odot. \eqno{(1)}$$ 
The large increase in pressure at the Galactic center suggests that
the typical self-gravitating mass would significantly decrease at the
center.  However, since the molecular clouds are known to be both
hotter and more turbulent at the center (G\"usten 1989), the strong dependence
of the Jeans mass on the velocity dispersion of the gas can be more
than offset by the increase in pressure.  The largest uncertainly in
applying Equation (1), is knowing what to take for $v_{\rm rms}$ for
the individual star forming cores.

	Astonishingly, even though the gas pressure is apparently
quite high, the rate of star formation in the nuclear region (which
can only be estimated for the massive stars) does not
seem to be significantly different from that of the Galactic disk
(G\"usten 1989).  This is remarkable
given that the surface pressure on the molecular clouds is
nearly three orders of magnitude greater than locally.  If true, it
suggests that the star forming properties of a GMC may be determined only
locally within a cloud and that the ambient conditions may have to be
even more extreme than those at the Galactic Center to have a 
significant effect on the IMF.  If so, it will be necessary 
to look at the internal structure of a GMC to see how it
organizes itself to form stars.  We turn to that question in the
following section.

\section{CLOUD STRUCTURE}
\subsection{Categorization}

Although molecular clouds are, by definition, regions in which the
gas is primarily molecular by {\it mass}, much of the {\it volume}
of such a cloud is not molecular. That is, the filling fraction of
molecular gas is low, $\simlt 20\%$ (Blitz 1993), and the cloud is highly
structured with large density variations from one location to another.
The structure (in volume density, column density, and velocity) of
molecular clouds reflects the conditions from which they form,
acts as a signpost to their evolution, and may, at sufficiently high
densities, be related to the mass scale of stars and the slope of the IMF.

\begin{figure}[htb]
\centerline{\psfig{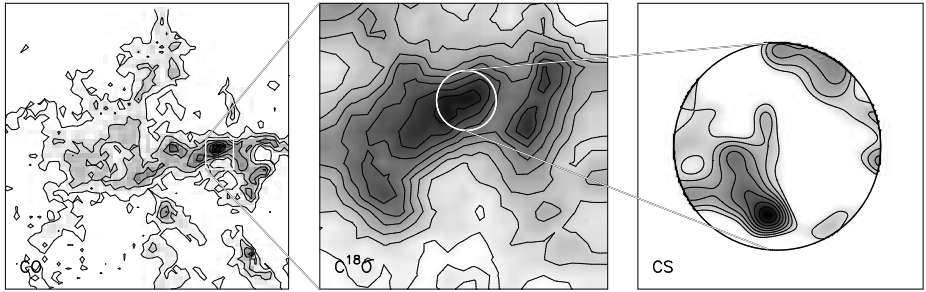}}
\caption{Hierarchical cloud structure. The three panels show a representative
view from cloud to clump to core. The bulk of the molecular gas (cloud; left
panel) is best seen in CO which, although optically thick, faithfully outlines
the location of the \h2. Internal structure (clumps; middle panel) is observed
at higher resolution in an optically thin line such as C$^{18}$O. With a
higher density tracer such as CS, cores (right panel) stand out.
The observations here are of the Rosette molecular cloud and are respectively,
Bell Labs ($90''$), FCRAO data ($50''$), and BIMA data ($10''$).}
\end{figure}

In this section, we discuss techniques to analyze cloud structure
and the results and implications of such analyses. First, we define
an operational categorization into clouds, clumps, and cores.
The scales of interest are illustrated in Figure 4.
This categorization is not inconsistent with the 
fractal models for cloud structure that are discussed in \S 3.4, 
although we argue that gravity introduces scales that limit
the range of validity of the fractal description.

Clouds are regions in which the gas is primarily molecular as stated above.
Almost all known molecular clouds in the Galaxy are detectable in CO.
Giant molecular clouds, with masses $\simgt 10^4~M_\odot$,
are generally gravitationally bound, and may contain several sites of
star formation. However, there are also many small molecular clouds
with masses $\simlt 10^2~M_\odot$, such as the unbound high latitude
clouds discovered by Blitz, Magnani, \& Mundy (1984), the chaff
discussed in \S 2.2, and the small gravitationally bound molecular
clouds in the Galactic plane cataloged by Clemens \& Barvainis (1988).
A small number of low mass stars are observed to form in some of these
clouds but their contribution to the total star formation rate in the
Galaxy is negligible (Magnani et al. 1995).  Heyer et al. (1999)
suggest that most of the clouds with M $< 10^3$ \msune are not
self-gravitating; the Clemens \& Barvainis clouds are presumably a
small but unknown fraction of mass of the low mass chaff.  

Clumps are coherent regions in $l-b-v$ space, generally identified
from spectral line maps of molecular emission.
Star-forming clumps are the massive clumps out of which stellar clusters form.
Although most clusters are unbound, the gas out of which
they form is bound (Williams et al. 1995).
Clumps may be blended together at low intensities, particularly
in low density molecular tracers such as CO and its isotopes.
In this case, several techniques exist to decompose the emission
into its constituent clumps (\S3.2).

Cores are regions out of which single stars (or multiple systems such as
binaries) form and are necessarily gravitationally bound. 
Not all material that goes into forming a star must come from
the core; some may be accreted from the surrounding clump
or cloud as the protostar moves through it (Bonnell et al. 1997).

\subsection{Structure analysis techniques}

Molecular cloud structure can be mapped via radio spectroscopy of
molecular lines (e.g., Bally et al. 1987),
continuum emission from dust (e.g., Wood, Myers, \& Daugherty 1994),
or stellar absorption by dust (Lada et al. 1994).
The first gives kinematical as well as spatial information and
results in a three dimensional cube of data, whereas the latter two
result in two dimensional datasets.
Many different techniques have been developed to analyze these data
which we discuss briefly here.

Stutzki \& G\"{u}sten (1990) and Williams, de Geus, \& Blitz (1994)
use the most direct approach and decompose the data into
a set of discrete clumps,
the first based on recursive tri-axial gaussian fits,
and the latter by identifying peaks of emission and
then tracing contours to lower levels.
The resulting clumps can be considered to be the ``building blocks''
of the cloud and may be analyzed in any number of ways to
determine a size-linewidth relation, mass spectrum, and variations
in cloud conditions as a function of position (Williams et al. 1995).
There are caveats associated with each method of clump deconvolution,
however. Since the structures in a spectral line map of a molecular cloud
are not, in general, gaussian, the recursive fitting method of
Stutzki \& G\"{u}sten (1990) will tend to find and
subsequently fit residuals around each clump, which
results in a mass spectrum that is steeper than
the true distribution. On the other hand, the contour tracing
method of Williams et al. (1994) has a tendency to blend small
features with larger structures and results in a mass spectrum
that is flatter than the true distribution.

Heyer \& Schloerb (1997) use
principal component analysis to identify differences
in line profiles over a map. A series of eigenvectors and
eigenimages are created which identify ever smaller velocity
fluctuations and their spatial distribution, resulting in the
determination of a size-linewidth relation.
Langer, Wilson, \& Anderson (1993) use Laplacian pyramid transforms
(a generalization of the Fourier transform) to measure the power on
different size scales in a map; as an application, they
determine the mass spectrum in the B5 molecular cloud.
Recently, Stutzki et al. (1998) have described a closely related
Allan-variance technique to characterize the fractal structure of
2-dimensional maps.
Houlahan \& Scalo (1992) define an algorithm that constructs
a structure tree for a map; this retains the spatial relation
of the individual components within the map but loses information
regarding their shapes and sizes. It is most useful for displaying
and ordering the hierarchical nature of the structures in a cloud.

Adams (1992) discusses a topological approach to quantify the
difference between maps. Various ``output functions''
(e.g., distribution of density, volume, and number of components
as a function of column density; see Wiseman \& Adams 1994)
are calculated for each cloud dataset and a suitably defined
metric is used to determine the distance between these
functions and therefore to quantify how similar clouds are,
or to rank a set of clouds.

A completely different technique was pioneered by Lada et al. (1994).
They determine a dust column density in the dark cloud IC\,5146
by star counts in the near-infrared and mapped cloud structure over a
much greater dynamic range ($A_V=0-32$~mag) than a single spectral line map.
The effective resolution, $\simgt 30''$, is determined by the sensitivity of
the observations.

The most striking result of applying these various analysis tools
to molecular cloud datasets is the identification of self-similar
structures characterized by power law relationships between,
most famously, the size and linewidth of features (Larson 1981),
and the number of objects of a given mass (e.g., Loren 1989).
Indeed, mass spectra are observed to follow a power law with
nearly the same exponent, $x=0.6-0.8$, where $dN/d\ln M \propto M^{-x}$
from clouds with masses up to $10^5~M_\odot$ in the outer Galaxy
to features in nearby high-latitude clouds with masses
as small as $10^{-4}~M_\odot$
(Heyer \& Terebey 1998; Kramer et al. 1998a; Heithausen et al. 1998).
Since a power law does not have a characteristic scale, the
implication is that clouds and their internal structure are scale-free.
This is a powerful motivation for a fractal description of the
molecular ISM (Falgarone et al. 1991, Elmegreen 1997a).
On the other hand, molecular cloud maps do have clearly identifiable
features, especially in spectral line maps when a velocity axis can be
used to separate kinematically distinct features along a line of sight
(Blitz 1993). These features are commonly called clumps,
but there are also filaments (e.g., Nagahama et al. 1998),
and rings, cavities, and shells (e.g., Carpenter et al. 1995).
Both the discrete (clump) and fractal description of clouds
can be used as tools of analysis and both reveal much about
cloud physics and star formation. We discuss each in turn in
the following subsections.

\subsection{Clumps}

Clump decomposition methods such as those described above by
Stutzki \& G\"{u}sten (1990) and Williams et al. (1994)
can be readily visualized and have an appealing simplicity.
In addition, as for all automated techniques, these algorithms
offer an unbiased way to analyze datasets, and are still a valid
and useful tool for cloud comparisons even if one does not subscribe
to the notion of clumps within clouds as a physical reality (Scalo 1990).

In a comparative study of two clouds, Williams et al. (1994)
searched for differences in cloud structure between star forming
and non-star forming GMCs.
The datasets they analyzed were maps of \thirteenco(1--0) emission
with similar spatial (0.7~pc) and velocity resolution (0.68~\kms)
but the two clouds, although of similar mass $\sim 10^5~M_\odot$,
have very different levels of star formation activity. The first, the
Rosette molecular cloud, is associated with an HII region
powered by a cluster of 17 O and B stars and also contains a number
of bright infrared sources from ongoing star formation deeper
within the cloud (Cox, Deharveng, \& Leene 1990).
The second cloud, G216-2.5, originally discovered by
Maddalena \& Thaddeus (1985),
contains no IRAS sources from sites of embedded star formation
and has an exceptionally low far-infrared luminosity to mass ratio 
(Blitz 1990),
$L_{\rm IR}/M_{\rm cloud}<0.07~L_\odot/M_\odot$, compared to
more typical values of order unity (see Williams \& Blitz 1998).

\begin{figure}[htb]
\centerline{\psfig{figure=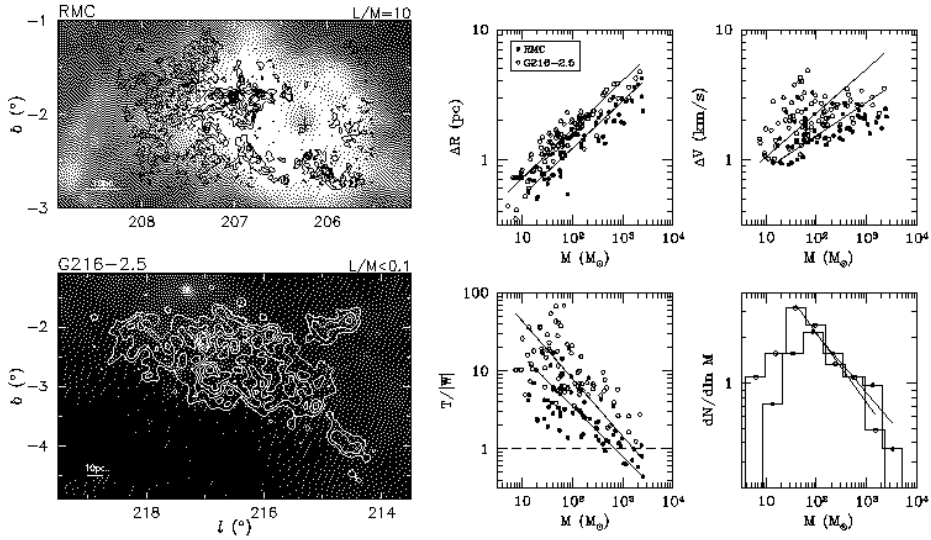,height=2.9in,angle=90,silent=1}}
\caption{Structure in the Rosette and G216-2.5 molecular clouds.
The two left panels show contours of velocity integrated CO emission
(levels at 15~K~\kms\ for the RMC, 1.8~K~\kms\ for G216-2.5)
overlaid on a grayscale image of the IRAS $100~\mu$m intensity
(1.1 to 2.5~MJy~sr$^{-1}$, same for both clouds).
The Rosette cloud is infrared bright, indicative of its high star
formation rate, but G216-2.5 has a very low infrared luminosity due
to a lack of star formation within it.
The four rightmost panels show power law relations between clump mass and
size, linewidth, energy balance (i.e. the ratio of kinetic energy,
$\calt=3M(\Delta v/2.355)^2/2$, to gravitational potential energy,
approximated as $\calw=-3GM^2/5R$),
and number (i.e. clump mass spectrum) for the two clouds.
The solid circles represent clumps in the RMC,
and open circles represent clumps in G216-2.5.
Each relationship has been fit by a power law: note that the power law
exponent is approximately the same for the clumps in each cloud despite
the large difference in star formation activity.}
\end{figure}

Almost 100 clumps were cataloged in each cloud, and sizes, linewidths,
and masses were calculated for each. These basic quantities were found to be
related by power laws with the same index for the two clouds,
but with different offsets (Figure~5) in the sense that for a given mass,
clumps in the non-star forming cloud are larger, and have greater linewidths
than in the star forming cloud.
The similarity of the power law indices suggests that,
on these scales, $\sim$~few~pc,
and at the low average densities, $\langle\nhtwo\rangle\sim 300$~\cc,
of the observed clumps, the principal difference between the star forming
and non-star forming cloud is the change of scale rather than the
collective nature of the structures in each cloud.

Figure~5 shows that the kinetic energy of each clump in
G216-2.5 exceeds its gravitational potential energy, and
therefore no clump in the cloud is self-gravitating (although the
cloud as a whole is bound).
On the other hand, Williams et al. (1995) show that,
for the Rosette molecular cloud, star formation occurs
only in the gravitationally bound clumps in the cloud.
Therefore, the lack of bound clumps in G216-2.5 may explain why
there is little or no star formation currently taking place within it.

Even in the Rosette cloud, most clumps are not gravitationally bound
(and do not form stars).
These unbound clumps have similar density profiles, $n(r)\propto 1/r^2$,
as the bound clumps (Williams et al. 1995), but contain relatively
little dense gas as traced by CO(3--2) or CS(2--1)
(Williams \& Blitz 1998).
The unbound clumps are ``pressure confined" in that their internal
kinetic pressure, which is primarily turbulent, is comparable to the
mean pressure of the ambient GMC (Blitz 1991; Bertoldi \& McKee 1992).
Simulations suggest that these clumps are transient
structures (Ostriker, Gammie, \& Stone 1999).

The nature of the interclump medium remains unclear:
Blitz \& Stark (1986) found low intensity broad line wings
in their CO and $^{13}$CO(1--0) observations of the Rosette molecular cloud
which, they postulated, resulted from a pervasive low density
molecular interclump medium, but Schneider et al. (1996) 
found that the line wings are also apparent in the $J=3-2$ transitions
with a line ratio similar to the core emission possibly indicating
a density not much different from the bulk of the cloud,
and therefore not interclump. On the other hand,
substantial HI envelopes exist around molecular clouds (Blitz 1990, 1993)
and may also pervade the volume between the clumps within a cloud
(Figure 6).
The turbulent pressures of the atomic and molecular components,
$\rho\sigma^2$, where $\rho$ is the density and $\sigma$ is the
velocity dispersion, are comparable (Williams et al. 1995)
and show that the HI can indeed confine the CO clumps.

\begin{figure}[htb]
\centerline{\psfig{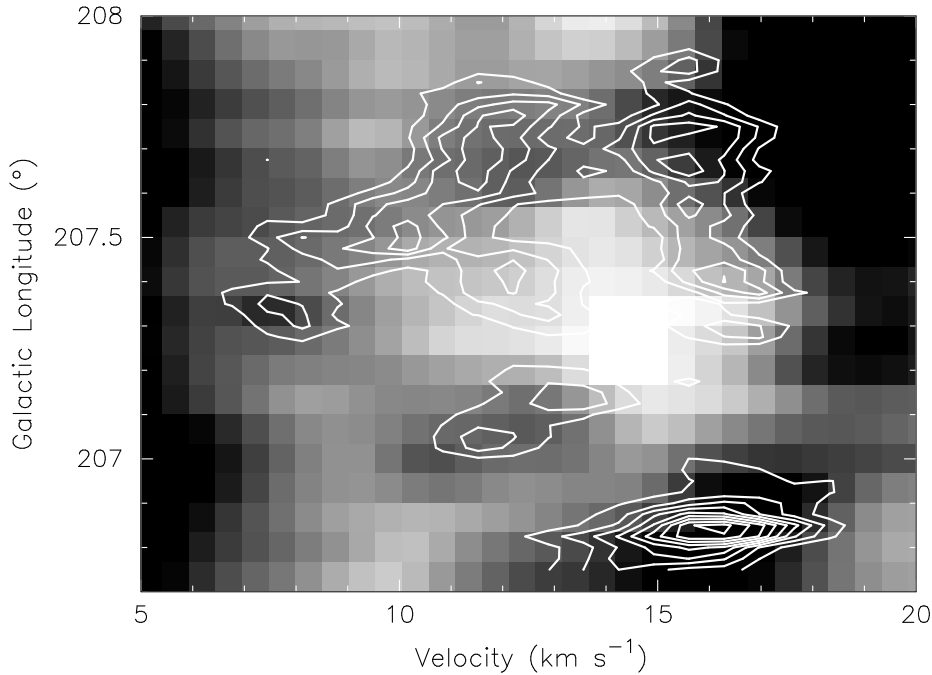}}
\caption{Detailed anti-correlation between HI and \thirteenco\ in
the Rosette molecular cloud. The grayscale (range 5 to 7~K~$^\circ$)
is HI emission (Kuchar \& Bania 1993),
and the contours (starting level and increment is 0.06~K~$^\circ$)
are \thirteenco\ data from Blitz \& Stark (1986). Each are a slice in
latitude integrated over $b=-2\fdg 1$ to $-1\fdg 9$.
The \thirteenco\ map shows a number of clumps, most of which are
gravitationally unbound, but they tend to lie in valleys in the
HI which may act as a pressure-confining medium.}
\end{figure}

Cloud, clump, and core density profiles are reflections of the
physics that shape their evolution, but the density profiles of clouds
and clumps have received scant attention.  For clouds, which often
are quite amorphous without a clear central peak, the density profile
is often difficult to define observationally.  For clumps, Williams
et al. (1995) showed that surface density profiles of pressure bound,
gravitationally bound, and star-forming clumps all have similar power
law indices close to 1.  Formally, the
fits range from -0.8 to -1.2, but these differences appear to
be only marginally 
significant.  For a spherical cloud of infinite extent, $\Sigma(r)\propto
r^{-1}$ implies $\rho(r) \propto r^{-2}$, suggesting that the
(turbulent) pressure support is spatially constant. However, McLaughlin \&
Pudritz (1996) argued that for finite spheres, the volume density
distribution can be considerably flatter than that inferred for
infinite clumps. Density distributions inferred from observations
also require consideration of beam-convolution effects.  It is
nevertheless astonishing that both strongly self-gravitating clumps
and those bound by external pressure have such similar, perhaps
identical density distributions.  Why this should be so is unclear.

\subsection{Fractal Structures}

An alternate description of the ISM is based on fractals.
High spatial dynamic range observations of molecular
clouds, whether by millimeter spectroscopy (e.g., Falgarone et al. 1998),
IRAS (Bazell \& D\'{e}sert 1988), or using the Hubble Space Telescope
(e.g., O'Dell \& Wong 1996; Hester et al. 1996)
show exceedingly complex patterns that appear to defy a
simple description in terms of clouds, clumps and cores;
Scalo (1990) has argued that such loaded names arose from
lower dynamic range observations and a general human tendency to
categorize continuous forms into discrete units.

As we have discussed above, it seems that however one analyzes
a molecular cloud dataset, one finds self-similar structures.
Moreover, the highly supersonic linewidths that are observed in
molecular clouds probably imply turbulent motions
(see discussion in Falgarone \& Phillips 1990), for which one
would naturally expect a fractal structure (Mandelbrot 1982).

The fractal dimension of a cloud boundary, $D$,
can be determined from the perimeter-area relation of a map,
$P\propto A^{D/2}$. Many studies of the molecular ISM
find a similar dimension, $D\simeq 1.4$ (Falgarone et al. 1991 and
references therein). In the absence of noise, $D>1$ demonstrates that cloud
boundaries are fractal. 
That $D$ is invariant from cloud to cloud
(star-forming or quiescent, gravitationally bound or not)
is perhaps related to the similarity in the
mass spectrum index in many different molecular clouds
(Kramer et al. 1998a; Heithausen et al. 1998).
Fractal models have been used to explain both the observed mass spectrum of
structures (Elmegreen \& Falgarone 1996) and the stellar IMF (Elmegreen 1997b).

Conclusions about physical processes in molecular clouds
that are drawn from the perimeter-size relation should be treated
with caution, however, since in
column density maps such as from IRAS observations or integrated
intensity spectral line maps, the observed structures are projections
of an inherently three-dimensional distribution which, for sufficiently
high filling factor, results in multiple overlapping of unrelated objects.
Such overlapping is found to mimic the observed fractal perimeter-size
relation in Monte-Carlo simulations of clumpy media by Witt \& Gordon (1996).
That is, in an inhomogeneous cloud the overlap of discrete objects
from projection can be a fractal even if the individual objects are not.
Furthermore, noise can produce a
non-integral exponent in the perimeter-area relation giving the
appearance of a fractal, but the degree to which noise affects the
published results has not yet been investigated. 

Probability density functions (PDFs) may be used to describe the distribution
of physical quantities (such as density and velocity) in a region of space
without resorting to concepts of discrete objects
such as clouds, clumps and cores.
Falgarone \& Phillips (1990), for example, have analyzed the PDF
of the velocity field of several clouds at different scales.
The low-level, broad line wings that are observed in
non-star forming regions show that the probability of rare, high-velocity
motions in the gas are greater than predicted by a normal (gaussian)
probability distribution. This {\it intermittent} behavior is expected
in a turbulent medium, and the detailed analysis of
Falgarone \& Phillips shows that the
deviations from the predictions for Kolmogorov turbulence are small.
Miesch \& Scalo (1995) calculate velocity centroid
PDFs from \thirteenco\ observations of star-forming regions and
also report non-gaussian behavior. Lis et al. (1996) compare their
results with a similar analysis applied to simulations of compressible
turbulence; such work may be a promising avenue for exploring the
role of turbulence in molecular clouds.  

Detailed simulations of 3-D hydromagnetic turbulence including gravity
have now become possible and may ultimately help to determine how the
structure in molecular clouds forms.  The best simulations to date are
those by Ostriker et al. (1999) and the next few years should show a
great deal of progress in the application of codes to molecular
clouds.  The utility of the clump finding algorithms such as
Clumpfind (Williams et al. 1994) 
applied to both the simulations and the observations will
be a good test of how closely the simulations match reality.

\subsection{Departures from self-similarity}

The universal self-similarity that is observed in all types of cloud,
over a wide range in mass and star forming activity is remarkable,
but a consequence of this universality is that it does not differentiate
between clouds with different rates of star formation (or those that are
not forming stars at all) and therefore it cannot be expected to explain
the detailed processes by which a star forms.
Star formation must be preceded
by a departure from structural self-similarity.

There have long been suggestions
that the thermal Jeans mass gives a scale 
that determines the characteristic mass of stars (Larson 1985).
In order to determine whether such a scale is important in molecular clouds,
Blitz \& Williams (1997) 
examined how the structural properties of a large scale, high resolution
\thirteenco\ map of the Taurus molecular cloud obtained
by Mizuno et al. (1995) varied as the resolution was degraded by
an order of magnitude. In their work, they use the temperature PDF
of the dataset to compare the cloud properties as a function of resolution.
This is the most basic statistic and requires minimal interpretation
of the data. In Figure 7 we show the temperature PDF
for the Taurus dataset at two resolutions
and four other \thirteenco\ maps of molecular clouds.

\begin{figure}[htb]
\centerline{\psfig{figure=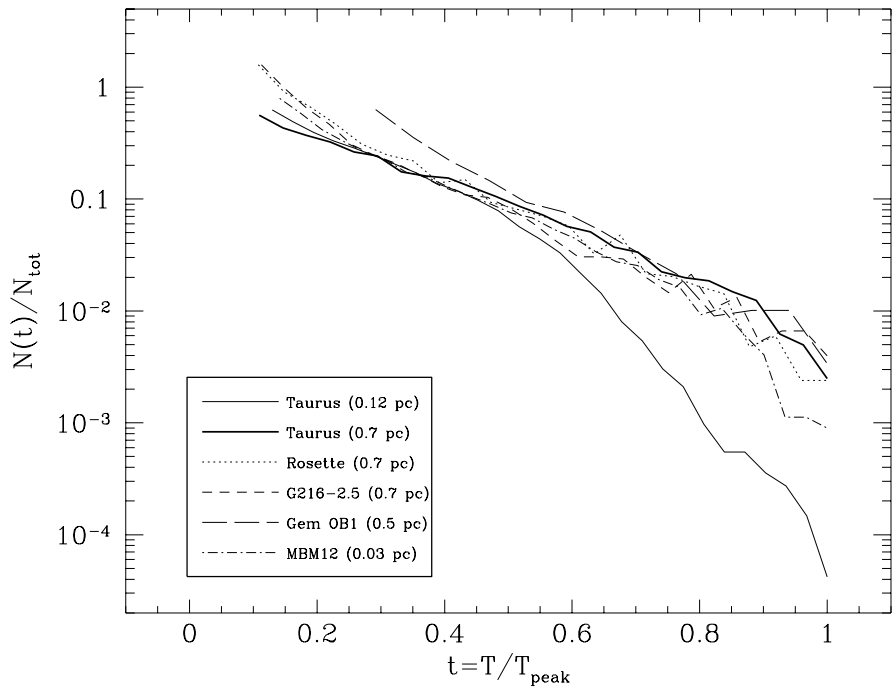,height=3.9in,angle=-90,silent=1}}
\caption{Intensity PDF in five molecular clouds,
all from \thirteenco(1--0) maps, except Gem OB1 which is CO(1--0).
Temperatures have been normalized by the peak value in each
dataset and then binned. The fraction of data points
with $T/T_{\rm peak}>0.3$
is plotted as a function of normalized temperature.
Note the similarity between the
different datasets, except for the high resolution Taurus map
which has a lower fraction of data points with temperatures
within 60\% of the peak.}
\end{figure}

To compare the different cloud PDFs, Figure 7 shows temperatures
that have been normalized by the peak,
$T_{\rm peak}$, of each map.
Each PDF has also been truncated at $T/T_{\rm peak}\simeq 0.15-0.25$
to show only those points with high signal-to-noise.
Within the Poisson errors (not shown for clarity), the PDFs of
the different clouds are all the same, except for the higher
resolution Taurus PDF for which there is a lower relative
probability of having lines of sight with $T/T_{\rm peak}\simgt 0.7$.
The low intensities of the \thirteenco\ emission imply that the optical
depth is small along all lines of sight in the map, and is not responsible
for this effect. Rather, from examination of the integrated intensity maps,
Blitz \& Williams (1997) show that this is due to a steepening of the
column density profiles at small size scales (see also Abergel et al. 1994).

There are two immediate implications from Figure 7.
First, the common exponential shape for the temperature PDF
is another manifestation of the self similar nature of cloud structure.
It is a simple quantity to calculate and may provide a quick and useful
test of the fidelity of cloud simulations.
Second, since the behavior of the Taurus dataset changes as it is
smoothed, it cannot be described by a single fractal dimension over
all scales represented in the map since the intensity PDF for a fractal
is invariant under smoothing (Figure~8).
There is other evidence for departures from self-similarity at 
similar size scales. Goodman et al. (1998) examine in detail the nature of
the size-linewidth relation in dense cores as linewidths approach a constant,
slightly greater than thermal, value in a central ``coherent'' region
$\sim 0.1$~pc diameter (Myers 1983).
Also, Larson (1995) finds that the two-point angular correlation function
of T Tauri stars in Taurus departs from a single power law at a size scale
of 0.04~pc (see also Simon 1997).

\begin{figure}[htb]
\centerline{\psfig{figure=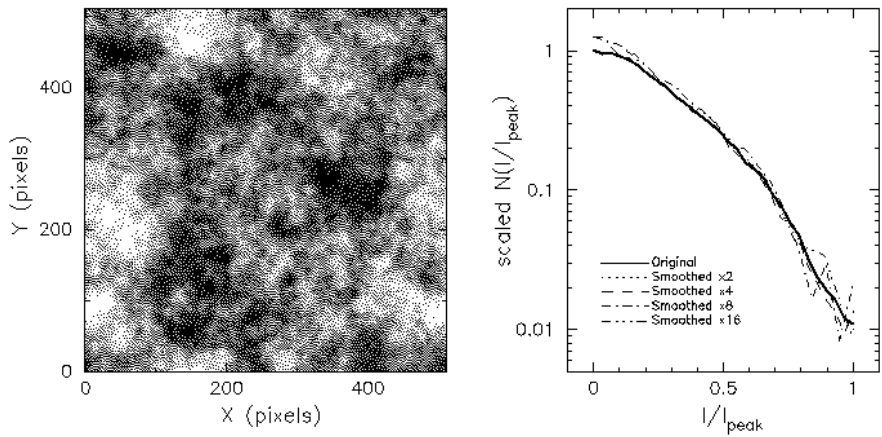,height=2.4in,angle=-90,silent=1}}
\caption{A fractal cloud and the effect of smoothing on its intensity PDF.
The left panel shows a grayscale image of a model fractal cloud from
$T=0$ to 1~K (courtesy of Chris Brunt).
The right panel shows the normalized temperature histogram for the original
dataset (heavy solid line) and then for the same dataset after binning
by 2, 4, 8, and 16 pixels. The intensity PDF is independent
of resolution as expected for this self-similar structure.}
\end{figure}

For gas of density $\nhtwo\sim 10^3$~\cc, these size scales
correspond to masses of order $\sim 1~M_\odot$, close to the
thermal Jeans mass at a temperature $T=10$~K. 
It is important to note that the
above evidence for characteristic scales comes from studies of
gravitationally bound, star forming regions:
self-similarity in unbound clouds continues to much smaller scales.
Figure 7 shows that the temperature PDF
of the unbound, high latitude cloud MBM12 is identical, at a
resolution of 0.03~pc, to the other lower resolution PDFs
of star forming GMCs. Similarly, the mass spectra of other
high latitude clouds follow power laws,
$dN/d\ln M \propto M^{-x}$ with $x\simeq 0.6-0.8$,
down to extremely low masses, $M\simeq 10^{-4}~M_\odot$
(Kramer et al. 1998a; Heithausen et al. 1998).
It appears to be the action of gravity that creates the
observed departures from self-similarity.
Note, however, that if the Jeans mass is physically relevant to the
structure of molecular clouds, then clouds must be magnetically supercritical
(i.e., magnetic fields cannot prevent dynamical collapse) since,
as explained in the chapter by Shu et al., the Jeans mass has
no meaning in subcritical clouds (such clouds are supported by magnetic
fields against gravitational collapse no matter how highly compressed).

\section{THE RELATION BETWEEN CLOUD STRUCTURE AND THE IMF}

The spectrum, lifetime, and end state of a star are primarily determined
by its mass. Consequently, the problem of understanding how the mass
of a star is determined during its formation, and the origin of the IMF,
has a very wide application in many fields from
galaxy evolution to the habitability of extrasolar planets. 
The form of the IMF is typically assumed to be invariant,
but since it is directly measureable only locally, knowing how it comes
about can help us predict how it might vary under different astrophysical
environments.

Many explanations for the form of the IMF use as their starting point
the mass spectrum of clouds and clumps as revealed by molecular line emission, 
$dN/d\ln M\propto M^{-x}$ with $x\simeq 0.5$. Most such structures, however,
are not forming stars: the majority of stars form in clusters in a few of
the most massive clumps in a cloud. An understanding of the origin
of the IMF can only come about with a more complete picture of 
the formation of star-forming clumps and the fragmentation of these
clumps down to individual star forming cores.

The cores that form an individual star (or multiple stellar system)
have typical average densities
$\nhtwo\sim 10^5$~\cc\ and can be observed in high excitation lines
or transitions of molecules with large dipole moments (Benson \& Myers 1989),
or via dust continuum emission at millimeter and sub-millimeter
wavelengths (Kramer et al. 1998b).
Because of their high densities, the surface filling fraction of
cores is low, even in cluster forming environments.
Therefore searches for cores have generally followed signs of
star formation activity, e.g. IRAS emission, outflows, etc.
and there have been few unbiased searches (e.g. Myers \& Benson 1983).
However, increases in instrument speed have now made it possible to survey
millimeter continuum emission over relatively large areas of the sky.
There have been two very recent results in this regard,
the first by Motte, Andr\'{e}, \& Neri (1998)
at 1.3~mm using the array bolometer on the IRAM 30~m telescope,
and the second by Testi \& Sargent (1998) at 3~mm using
the OVRO interferometer.

Motte et al. (1998) mapped the $\rho$ Ophiuchus cloud,
the closest rich cluster forming region
and Testi \& Sargent (1998) mapped the Serpens molecular cloud,
a more distant, but richer star forming region.
In each case, the large scale, high resolution observations reveal
a large number of embedded young protostars and also starless,
dense condensations. Both studies find that the mass spectrum
of the cores is significantly steeper, 
$x>1.1$ (where $dN/d\ln M\propto M^{-x}$) than clump mass spectra,
$x\simeq 0.6-0.8$. The core mass spectra approach the slope, $x=1.35$, 
of the stellar (Salpeter) IMF.
These intriguing results suggest that the masses of stars that form
in a cloud are directly linked to its structure.
However, these investigations could not determine whether the clumps
are self-gravitating, a necessary condition to show that these
continuum cores actually form stars.  Recall, for example, that the
lowest mass molecular clouds, as well as the low mass clumps within a
GMC are not, in general, self-gravitating.
Surveys of other cluster forming regions should be made and these
studies should be followed up with spectral line observations
to determine whether the starless cores are self-gravitating,
or even collapsing (Williams \& Myers 1999),
and therefore likely progenitors of future stars.

It is worth noting also that, since the filling fraction of the cores is
much lower than that of the lower density CO clumps, the
projection effects that we cautioned about above and simulated by
Witt \& Gordon (1996) are much less severe.
Thus, it is not entirely clear that the observed departure
from self-similarity is due to real physical processes such as gravity
and/or the dissipation of turbulence, or is simply a result of the lower
filling fraction.

As high resolution studies of individual cores
in cluster environments become more commonplace, the relationship between
stellar mass and core mass will be determined more precisely.
If the core mass spectrum is indeed similar to the stellar IMF,
then the fraction of a cores mass that goes into a star
(the star formation efficiency of the core)
must be approximately independent of mass and the stellar IMF
is determined principally by the cloud fragmentation processes.
By measuring the core mass spectrum in different clusters in a variety
of different molecular clouds, the influence of the large scale structure
and environment on the IMF can be quantified.

\section{Summary}

Considerable progress has been made in addressing many of the issues
concerning molecular clouds since the first Crete conference and
new results tentatively 
suggest a direct link between cloud structure and the IMF.

It seems clear that molecular clouds form through the condensation of HI
since very little molecular gas exists between the spiral arms.
Collisional agglomeration in any form appears to be ruled out because of
the small total mass of low mass clouds (the chaff).
However, many details of the formation process, such as the
role of Galactic shocks and magnetic fields for instance,
remain to be understood.

A close association between atomic and molecular gas is evident in the
solar neighborhood and detailed maps suggest that the interclump medium
within clouds is predominantly atomic.
In the Galactic Center, however, the formation of molecular clouds
and their association with the atomic gas must be very different because
the pressure is more than 2 orders of magnitude higher, and the atomic mass
fraction more than 2 orders of magnitude lower, than at the solar circle.
Understanding cloud and star formation in this environment is a step
toward understanding how stars form in even more extreme environments
such as starbursts.

The structure within clouds reflects their formation from an inhomogeneous
atomic ISM and, at moderate densities, $\nhtwo\simlt 10^3$~\cc,
is self-similar up to a scale set by self-gravity.
In this regime, scaling laws such as clump mass spectra have similar
power law indices independent of the star-forming nature of the cloud.

At higher densities, and smaller sizes, as linewidths approach their
thermal value, structures depart from the same self-similar description.
This departure may mark the boundary between cloud evolution and
star formation. Clusters of individual star forming cores,
with a mass spectrum that approaches the Salpeter IMF,
are observed in the $\rho$ Ophiuchus and Serpens clouds.
The study of the structure, dynamics, and distribution of these
cores will lead to a better understanding of the
relationship between the structure and evolution
of molecular clouds and the initial mass function of stars.

\vskip 0.2in
Funding has been provided by a grant from the NSF.
JPW is supported by a Jansky fellowship.
We thank Tamara Helfer for providing Figure~1, 
Mark Heyer for Figure 2, and Loris Magnani for
results prior to publication.  We would also like to acknowledge
discussions with Mark Heyer, Tam Helfer, Chris McKee, Tom Sodroski and
Michele Thornley, about various topics covered in this review.


\end{document}